\documentclass[pra,aps,floatfix,superscriptaddress,twocolumn,reprint]{revtex4-2}
\usepackage{graphicx} % Required for inserting images
\usepackage{physics}
\usepackage{cancel,amsmath,amsfonts,amssymb}
\usepackage{mathtools,comment}
\usepackage{bm}
\usepackage{color}

\usepackage{csquotes}
\usepackage[toc]{appendix} % anche in indice
\usepackage{amsthm}

\usepackage{amssymb}

\usepackage{hyperref}
\hypersetup{
  colorlinks=true, % Attiva i colori dei collegamenti
  linktoc=page,    % Rendi cliccabili solo i numeri delle pagine
  linkcolor=blue,  % Colore dei collegamenti
  citecolor=blue,  % Colore dei collegamenti per le citazioni
  urlcolor=blue,   % Colore dei collegamenti URL
}

\definecolor{darkgreen}{rgb}{0,0.5,0}

\newcommand{\lap}{L}
\graphicspath{{./Immagini/}}
\usepackage{subcaption}
\usepackage{float}

\begin{document}
\title{Temporal nonclassicality in continuous-time quantum walks}
\author{Paolo Luppi}
\email{paolo.luppi@unimi.it}
\affiliation{Dipartimento di Fisica ``Aldo Pontremoli'', Universit\`a degli Studi di Milano, via~Celoria~16, I-20133 Milan, Italy}
\affiliation{INFN, Sezione di Milano, Via Celoria 16, I-20133 Milan, Italy}
\author{ Claudia Benedetti}
\email{claudia.benedetti@unimi.it}
\affiliation{Dipartimento di Fisica ``Aldo Pontremoli'', Universit\`a degli Studi di Milano, via~Celoria~16, I-20133 Milan, Italy}
\affiliation{INFN, Sezione di Milano, Via Celoria 16, I-20133 Milan, Italy}
\author{ Andrea Smirne }
\email{andrea.smirne@unimi.it}
\affiliation{Dipartimento di Fisica ``Aldo Pontremoli'', Universit\`a degli Studi di Milano, via~Celoria~16, I-20133 Milan, Italy}
\affiliation{INFN, Sezione di Milano, Via Celoria 16, I-20133 Milan, Italy}
\date{\today}

\begin{abstract}
Quantum walks represent a paradigmatic framework to explore and manipulate quantum behaviors. 
In this paper, we investigate the genuinely quantum features of continuous-time quantum walks by combining a single-time and a multi-time
quantifier of nonclassicality. On the one hand, we consider the quantum--classical dynamical distance $D_{\mathrm{QC}}(t)$, which measures the departure of the time-evolved quantum state
of a continuous-time quantum walk from the classical state of a random walk on the same graph. 
On the other, we analyse the joint probability distributions associated with sequential measurements of the walker's position,
assessing their violation of the classical Kolmogorov consistency conditions
via a dedicated quantifier $\bar{K}(t)$.    
We demonstrate a quadratic short-time scaling of $\bar{K}(t)$, which differs from the known linear scaling of $D_{\mathrm{QC}}(t)$, but, as the latter, is fully determined by the degree of the initially occupied node and is independent of the global graph topology. At longer times, instead, $\bar{K}(t)$ exhibits a pronounced topology-driven behavior: 
it is strongly suppressed on complete graphs while remaining finite and oscillatory on cycles, in contrast with the almost topology-independent asymptotics of $D_{\mathrm{QC}}(t)$. We then extend the analysis to Markovian open-system dynamics, focusing on dephasing in the position basis (Haken–Strobl model) and in the energy basis (intrinsic decoherence). Site dephasing drives both quantifiers to zero, with the decay of $\bar{K}(t)$ controlled by the spectral gap of the corresponding Lindblad generator. By contrast, energy-basis dephasing preserves a finite asymptotic value of $\bar{K}(t)$, 
depending on the overlap structure of the Laplacian eigenspaces with the site basis.
%contrary to $D_{\mathrm{QC}}(t)$, even in the absence of spectral degeneracy of the Laplacian. 
Our results demonstrate that distinct notions of nonclassicality lead to qualitatively and quantitatively different assessments of how quantum a given walk is, pointing at the intricate nature of temporal quantum correlations in networked quantum systems under realistic decoherence.
\end{abstract}
\maketitle

\section{Introduction}
Continuous-time quantum walks (CTQWs) are the quantum analogue  of classical continuous-time random walks \cite{fahri98,kempe003,Portugal,VenegasAndraca2012}.
Since their introduction, CTQWs have  found  application across  a variety of research areas, including universal quantum computation \cite{universalQC,qiang24,kendon20}, quantum algorithms \cite{quantumsearch,Apers2022,QWalg,Chakraborty20,candeloro22}, quantum transport \cite{mulken2011,mulken07,Mulken2007,tamascelli19,maciel20,finocchiaro25} and as probes of graph topologies \cite{gianani23,benedetti24,campbel25,Romeo25}. 
Experimentally, CTQWs have been realized in different  physical platforms, such as   waveguide arrays \cite{Perets2008,peruzzo10,Biggerstaff16,Caruso16,benedetti21}, trapped ions \cite{tamura20}, microwave \cite{bohm15} and photonic settings \cite{Wang:20,imany20}. Experimental realizations, together with the growing range of applications, have motivated  interest in identifying and quantifying the genuinely quantum features of CTQWs, as well as in understanding how such features manifest in realistic settings.
 
In this context, quantum behaviors are typically understood as distinctive features in the evolution of the walker’s position probability distribution 
-- a paradigmatic example being the doubly peaked distribution for a quantum walker on a line, which contrasts with the Gaussian distribution
of a classical walker \cite{konno2005,bessen2006,Mallick2019}. 
Along this line, the comparison of the evolutions of, respectively, the quantum state of a CTQW with a classical random-walk probability distribution is used not only
to discriminate between quantum and classical walks, but also to quantify their difference.
For instance, the
quantum--classical dynamical distance $D_{\mathrm{QC}}(t)$ measures how the quantum evolution deviates from a classical random walk on the same graph
via quantum fidelity~\cite{QCD}.

This kind of approach relies on the comparison between the quantum walker's state at a given time and the position probability distribution
of a specific classical walk, which is defined on the same graph and fixes a diagonal structure on the position basis;
however, it does not account for the comparison of the quantum walk with \emph{any} possible classical walk. 
In fact, it has been shown that, for various types of random walks, the population distribution generated by a quantum walk
can be reproduced by suitable classical walks, 
if one allows time-inhomogeneity or memory in the transition probabilities~\cite{deFalco2008,Montero2016,Montero2017,Andrade2020}.

From a broader perspective, quantum predictions can be unambiguously discriminated from any classical prediction 
by looking at multi-time probabilities, associated with sequential measurements at different times. 
This is the key idea behind the Leggett-Garg inequalities \cite{Leggett1985,Leggett2002,Emary2014,Halliwell2019,Vitagliano2023},
and it is based on the possibility, that is present only at the classical level, 
of accessing the value of any observable without disturbing it and without disturbing the value of any subsequently measured observable.
In other terms, classically, and only classically, measuring and not making any selection according to the measurement outcome
(more pictorially, measuring and forgetting the outcome) is the same as not measuring at all.
Within the context of sequential measurements at different times, this property of classical systems is expressed by the Kolmogorov consistency conditions -- i.e., the defining properties of classical stochastic processes \cite{Feller1971} -- 
whose violation then reveals genuinely quantum temporal behaviors \cite{Smirne2018,Strasberg2019,Milz2020,Lonigro2022,Strasberg2023,Szankowski2024,Budini2025}.
Indeed, among many different applications, the Leggett-Garg inequalities \cite{Robens2015} and the Kolmogorov consistency conditions \cite{Nitsche2018,Smirne2020} have been used to witness quantum behaviors also in (discrete-time) quantum walks.

In this work, we make an extended comparison of the signatures of nonclassicality in CTQWs witnessed by the 
quantum-classical dynamical distance $D_{\mathrm{QC}}(t)$ and the multi-time nonclassicality captured by the violation of the Kolmogorov
consistency conditions. The latter is quantified by examining the two-time 
joint probability distributions and by introducing a time-averaged Kolmogorov violation $\bar{K}(t)$, obtained by integrating over the intermediate measurement time.
Our analysis includes both closed- and open-system \cite{Breuer2002} dynamics of the walker. For the latter,  
environmental effects are modeled in a fully Markovian framework, where the dynamics is fixed by a Lindblad master equation \cite{Gorini1976,Lindblad1976} and the multi-time probabilities
by the quantum regression formula \cite{Lax1968,Swain1981,Carmichael1993,Breuer2002,Guarnieri2014}.
We find that the two quantifiers show quite different behaviors. At short times, they display distinct scaling laws: $D_{\mathrm{QC}}(t)$ grows linearly with time, while $\bar{K}(t)$ exhibits a quadratic scaling; yet both depend only on the degree of the initial node, independently of the global graph topology. For unitary walks, at long times the quantum--classical distance is known to converge to a topology-independent asymptotic value~\cite{QCD}, whereas our analysis reveals that $\bar{K}(t)$ exhibits a pronounced topology-driven behavior: it is strongly suppressed on complete graphs while remaining finite and oscillatory on cycles. Moving to the open-system scenario, 
we show that site-basis dephasing drives $\bar{K}(t)$ to zero, while energy-basis dephasing preserves a finite long-time value,
contrary to $D_{\mathrm{QC}}(t)$,
even in the absence of spectral degeneracy of the Laplacian that fixes the graph. 

The remainder of the paper is organized as follows. In Section \ref{sec:ctqw}, we briefly recall the theoretical framework of CTQWs and introduce the quantum--classical dynamical distance as a benchmark single-time quantifier. Section \ref{sec:mtn} presents the definition of multi-time probabilities for an open quantum system in the Markovian regime, as well as the Kolmogorov-based measure of temporal nonclassicality and its main properties. 
Section~\ref{sec:kni} provides a detailed analysis of nonclassicality in unitary quantum walks, emphasizing universal trends and topology-induced effects. 
Section~\ref{sec:oqw} extends the discussion to open quantum walks with both site- and energy-basis dephasing, 
focusing in particular on the presence or absence of residual nonclassicality in the long-time regime. 
Finally, Section~\ref{sec:cao} summarizes our main findings and outlines perspectives for future research.

\section{Unitary and open-system dynamics of continuous-time quantum walks}\label{sec:ctqw}

We start by fixing notation for CTQWs, graph Laplacians, and the open-system models used throughout the paper. 
These ingredients provide the starting point for both the single-time and multi-time nonclassicality introduced later.

\subsection{Unitary dynamics}
A CTQW  describes  the unitary evolution of a quantum particle  in a discrete position space.
The positions that the walker can occupy, together with their connections, are mathematically represented by a graph 
 $G(V,E)$, where $V$ and $E$  are the sets of vertices and edges respectively \cite{fahri98,kempe003,Portugal,VenegasAndraca2012}.
  The Hilbert space of the quantum walker is spanned by the basis   vectors $\{\ket{k}\}_{k=0}^{N-1}$ which correspond to   localized states on the $N$  sites of the graphs. 
 The dynamics of the walker is governed by a Hamiltonian $H$ that encodes the  topology of the underlying graph. A common choice for $H$  is the graph Laplacian $L$, such that the Hamiltonian is a real symmetric matrix: 
 \begin{align}
    H =  r L,
\end{align} 
where $ r > 0$ is the transition rate. The off-diagonal elements of the   Laplacian $L$ are $L_{jk} = -1$ if vertices $j$ and $k$ are connected and $L_{jk}=0$ otherwise, while the  diagonal entries are $L_{jj} = d_j$, with $d_j$ denoting  the degree of node $j$.
Graphs where each node has the same degree, $d_j=d$ for all $j=0,\dots,N-1$ are called regular graphs.

The parameter $r$ acts as a temporal scaling factor; 
in the following we set $ r = \hbar = 1$, making both time and energy dimensionless.
Moreover, we focus on two regular graph topologies, representatives of low and maximal connectivity: the cycle and the complete graph. 
The Laplacian matrices of both graphs are circulant \cite{olson14}, which enables an analytical treatment. For graphs with $N$ vertices, the Laplacian eigenvectors for both topologies are given by  the discrete Fourier modes:
\begin{equation}
    \ket{\psi_k} = \frac{1}{\sqrt{N}} \sum_{j=0}^{N-1} e^{i \frac{2\pi}{N}kj} \ket{j}, \qquad k=0,\dots,N-1.
    \label{eigenstatesCirc}
\end{equation}
These states correspond to delocalized wavefunctions  whose spatial probability distributions are uniform over all sites.

In the cycle graph  each node is connected to its two nearest neighbours, making it suitable to describe systems with nearest-neighbors interactions. The degree of each vertex is $d=2$,
thus making it the least connected topology among regular graphs.
Its Laplacian eigenvalues are:
\begin{align}
   \lambda_k^{\scriptsize \text{(cycle)}} = 2 - 2 \cos\!\left(\frac{2\pi k}{N}\right), \qquad k = 0,1,\dots,N-1.
   \label{lcycl}
\end{align}
In contrast,  in the complete graph  each vertex is connected to all other nodes, with vertex degree $d=N-1$. It has maximal connectivity, and it corresponds to systems with all-to-all interactions.
The Laplacian of the complete graph has two distinct eigenvalues: 
\begin{align}
    \lambda_k^{\scriptsize \text{(compl)}}=N(1-\delta_{k0}) \qquad k = 0,1,\dots,N-1,
\end{align} where $\delta_{k0}$ is the Kronecker delta. The zero eigenvalue corresponds to  the ground state while the eigenvalue $N$ corresponds to  the remaining $(N-1)$ orthogonal states that span the degenerate subspace.
The  walker evolves under the action of the unitary evolution operator $U(t)=e^{-iLt}$, so that, assuming a quantum walk initially localized at node $\ket{\nu}$, the  state at time $t$   is:
\begin{equation}
    \ket{\psi_\nu(t)} = \sum_k \alpha_{k\nu}(t)\ket{k},
    \label{unitQW}
\end{equation}
where the transition amplitudes between nodes are given by  $\alpha_{k\nu}(t) = \bra{k}e^{-iLt}\ket{\nu}$.
Eq.~\eqref{unitQW} describes the unitary evolution of a closed system. 

\subsection{Open-system dynamics}
When the walker interacts with  an external environment, the dynamics becomes non-unitary due to decoherence and dissipation effects.
In experimental implementations of quantum walks, such interactions with the environment are unavoidable. This raises the crucial question of whether, and to what extent, the open-system dynamics degrade the quantum characteristics of the evolution, so that a thorough understanding of decoherence effects in CTQWs is essential for their reliable use. 
%In particular, it enables the development of strategies to mitigate, control, or even exploit the presence of noise.

To deal with open-system dynamics, the state of the system has to be described by a density operator 
$\rho(t)$, whose time evolution is governed by a quantum master equation. 
We restrict our analysis to Markovian open-system dynamics, so that the time evolution of the density operator 
is fixed by the Lindblad master equation~\cite{Gorini1976,Lindblad1976,Breuer2002}:
\begin{equation}    \label{eq:Lindblad}
    \dot{\rho}(t)=\mathcal{L}[\rho(t) ] =  -i[\lap,\rho(t)] + \mathcal{D}[\rho(t)],
\end{equation}
where $\mathcal{D}$ is the dissipator defined by   
\begin{equation}\label{eq:diss}
    \mathcal{D}[\rho]= \sum_{k=1}^{N^2-1} \gamma_k \left( G_k \rho G_k^\dagger - \frac{1}{2} \{ G_k^\dagger G_k, \rho \} \right),
\end{equation}
which accounts for the decoherence and dissipation induced by the interaction with the environment.
The formal solution of Eq.(\ref{eq:Lindblad}) can be written as $\rho(t)=e^{\mathcal{L}t}[\rho(0)]$,
with $\rho(0)=\ketbra{\nu}$ the initial condition that we will fix throughout our analysis.
In Eq. \eqref{eq:Lindblad}, the coefficients $ \gamma_k \geq 0$ are positive rates and the $G_k$ are operators that 
describe the impact of the interaction with the environment on the open-system's dynamics.
In this work we focus on  two important, but different, decoherence mechanisms: decoherence 
%in the position basis 
described by the Haken-Strobl master equation \cite{Hakentight} and  intrinsic decoherence, \textit{i.e.}, corresponding to dephasing in the energy eigenbasis of the system Hamiltonian \cite{intrinsicdecoherence}.
\\
\\
The Haken-Strobl master equation describes the dynamics of a quantum system interacting   with a local environment that induces dephasing in the site basis. 
It is widely used to study  excitation transport in systems coupled to phonons, in   photosynthetic complexes and light-harvesting networks~\cite{Hakennetwork,gaab04,catalano23}.
In this model, environmental fluctuations cause random detunings of on-site energies that do not affect population transfer, but only coherences.
The Haken-Strobl operators are local site projectors 
$G_k = \ketbra{k}{k}$ for $k=0\dots N-1$. If we assume a uniform dephasing rate $\gamma_k=\gamma$ for all sites, 
 the Lindblad master equation in the node basis simplifies to:
\begin{equation}
   \dot{\rho}(t)=
     -i[\lap,\rho(t)] + \gamma \left(\sum_{k=0}^{N-1}  \rho_{kk}(t)|k\rangle\langle k|  -  \rho(t) \right).
    \label{eq:Haken}
\end{equation}
 Eq.~\eqref{eq:Haken} shows that the dephasing term 
 %exponentially 
 suppresses all the off-diagonal elements of the density matrix at rate $\gamma$, independently of the distance between sites, whereas the populations $\rho_{kk}(t)$ remain unaffected and evolve only under the coherent Hamiltonian dynamics.
\\
 The second  decoherence mechanism we consider  induces pure dephasing in the energy eigenbasis of the Hamiltonian. The corresponding master equation is obtained from the Lindblad  equation with a single Lindblad operator coinciding with the system Hamiltonian (Laplacian) and can be written as:
\begin{equation}
\dot{\rho}(t) = -i[\lap,\rho(t)] - \frac{\gamma}{2}[\lap,[\lap,\rho(t)]],
\label{eq:decoherenceEnergy}
\end{equation}
with $\gamma$ acting again as the decoherence rate. 
The solution of Eq.\eqref{eq:decoherenceEnergy} is given by \cite{DQCdecoherence}
\begin{equation}
\rho(t)=\sum_{\lambda,\lambda'} \Xi_{\lambda}\rho(0)\Xi_{\lambda'}
e^{-i(\lambda-\lambda')t-\frac{\gamma}{2}(\lambda-\lambda')^2 t},
\label{eq:soluzioneesattaenergy}
\end{equation}
where we introduced the eigenvalues of the Laplacian and the corresponding projectors into the eigenspaces:
given the eigenbasis $\{\ket{\psi_k}\}_{k=1,\ldots,N}$,
\begin{equation}
    L \ket{\psi_k} = \ell_k \ket{\psi_k},
\end{equation}
one has
\begin{equation}\label{eq:xil}
    \Xi_{\lambda} = \sum_{k : \ell_k =\lambda} \ket{\psi_k}\bra{\psi_k}.
\end{equation}
Populations in the energy eigenbasis are left unchanged by the evolution in Eq.(\ref{eq:soluzioneesattaenergy}), while coherences decay exponentially. For non-degenerate eigenstates, the decay rate increases with the energy gap $|\lambda-\lambda'|$.
This model is often referred to as  \emph{intrinsic decoherence}, and it was originally introduced to describe  stochastic quantum jumps in time \cite{intrinsicdecoherence} and  used in the context of 
qubit systems \cite{kimm02,alenezi22} and driven quantum systems~\cite{laserdecoherence}.\\

\subsection{Quantum-classical dynamical distance}\label{sec:qcdd}
The evolution of a QW on a given graph is fundamentally different from its classical counterpart, i.e.   a random walk. 
A continuous-time random walk describes the stochastic motion of a classical particle over a graph, characterized by the Laplacian  $L$,
which fixes the  transition probability from site $\nu$ to site $k$ at time $t$ according to
\begin{equation}\label{eq:toc}
p_{k\nu}(t) = \bra{k}e^{- L t}\ket{\nu}.
\end{equation}
In the quantum setting, the state of the classical walker at time $t$ can be represented by a diagonal density matrix, whose diagonal entries correspond to the classical transition probabilities:
\begin{align}
    \mathcal{E}_C(t)=  \sum_k p_{k\nu}(t)\ketbra{k}{k}.
\end{align}
Relying on this, a  single-time measure, called the quantum-classical dynamical distance $D_{\mathrm{QC}}(t)$, has been introduced to quantify the difference between classical and quantum  dynamics on the same graph \cite{QCD,QCDapp}:
it is defined as
\begin{equation}
    D_{\mathrm{QC}}(t) \equiv 1 - \min_{\rho_0} F\!\left[ \mathcal{E}_C(t),\,\rho(t)\right],
    \label{qcd}
\end{equation}
where $\rho_0$ is a localized initial state, $\rho(t)$ is the corresponding quantum state evolving according to Eq. \eqref{eq:Lindblad} and   $F(\rho_1,\rho_2) = \left[\mathrm{Tr}\sqrt{\sqrt{\rho_1}\rho_2\sqrt{\rho_1}}\,\right]^2$ is the quantum fidelity between two states.
Note that in regular graphs all nodes are equivalent, so that any localized initial state satisfies the minimization condition in  Eq. \eqref{qcd}. \\
 The quantity $ D_{\mathrm{QC}}(t)$ quantifies the deviation between the quantum and classical evolutions at time $t$: $D_{\mathrm{QC}}(t)=0$ only when the two dynamics yield indistinguishable probability distributions, indicating that the QW single-time probabilities coincide with those
 obtained for a classical walker whose evolution is ruled by the same Laplacian.
 In this sense, the quantum-classical distance can be used as an operational  measure of quantumness of the walker dynamics. 

\section{Multi-time Nonclassicality}\label{sec:mtn}

\begin{figure}[H]
    \centering
    \includegraphics[width=0.5\textwidth]{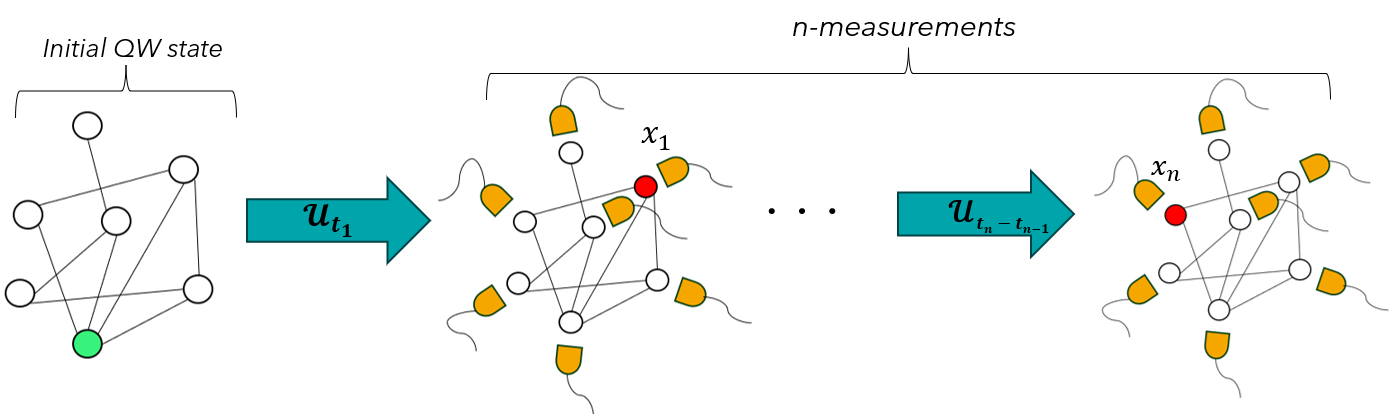}
    \caption{Schematic illustration of a multi-time measurement protocol: the quantum walker is sequentially measured at different times $t_1, t_2, \dots,t_n$, yielding joint probability distributions that can be used to quantify temporal nonclassicality.}
    \label{fig:schema}
\end{figure}

In the previous section, we focused on the single-time probability distributions of a CTQW and a notion of nonclassicality associated with them.  
Here we move to a multi-time perspective and consider the joint outcome statistics for sequences of measurements performed at different times,
assessing their departure from
a classical stochastic description, see Fig.\ref{fig:schema}.

\subsection{Kolmogorov-based quantifier of multi-time nonclassicality}

Again, we start with the case of a unitary evolution of the walker. Hence, consider a quantum walk initialized in the state $\rho(0)$, evolving unitarily under $U_t = e^{-iLt}$, and a (discrete-value) observable 
\begin{equation}
    X = \sum_x x \, \Pi_x ,
\end{equation}
where $\Pi_x$ is the projector onto the eigenspace associated with the outcome $x$. 
Performing projective measurements of $X$ at times $t_1 \leq t_2 \leq \dots \leq t_n$, the joint probability distribution 
of getting the sequence of outcomes $x_1, x_2, \dots, x_n$
is given by
\begin{equation}\label{eq:multi}
    P_n ( x_n,t_n;\dots;x_1,t_1 ) 
    = \mathrm{Tr}\!\left\{ \mathcal{M}_{x_n}\,\mathcal{U}_{t_n-t_{n-1}}\cdots 
    \mathcal{M}_{x_1}\,\mathcal{U}_{t_1}\rho(0)\right\},
\end{equation}
where we introduced the superoperators
\begin{equation}
    \mathcal{U}_t\rho = U_t \rho U_t^\dag, 
    \qquad 
    \mathcal{M}_x\rho = \Pi_x \rho \Pi_x ,
\end{equation}
implying that each superoperator acts on everything at its right.

Having defined the family of multi-time quantum probabilities, we now turn to the criterion that we will use to discriminate
them from multi-time classical probabilities, i.e., from the joint probability distributions associated with classical stochastic processes.
Such a criterion is provided by the Kolmogorov consistency conditions, which are (along with positivity and normalization) 
necessary and sufficient conditions that a family of joint probability distributions must satisfy in order to be reproduced by
an underlying classical stochastic process \cite{Feller1971}. 
Explicitly, given any $n$-time joint probability distribution 
$P_n(x_n,t_n; \dots; x_1,t_1)$ 
for any sequence of outcomes $\{x_n,\dots,x_1\}$ at any sequence of times $t_n\geq \dots \geq t_1$, the 
Kolmogorov consistency conditions are given by
\begin{eqnarray}
    &&P_{n-1}(x_n,t_n; \dots; \cancel{x_k,t_k}; \dots; x_1,t_1) \nonumber\\
    &&= \sum_{x_k} P_n(x_n,t_n; \dots; x_k,t_k; \dots; x_1,t_1),
\end{eqnarray}
that is, marginalization over the outcome -- i.e., measuring and not performing any selection according to the measurement outcome -- 
at any time $t_k$
is equivalent to not performing any measurement at that time. 
This reflects the classical intuition that in a stochastic process measurements simply reveal pre-existing properties of the system. 
In contrast, in quantum mechanics also non-selective measurements can change the state of the system, thereby disturbing the subsequent evolution and leading to a violation of the Kolmogorov consistency conditions. 

In this work we restrict our analysis to two-time probabilities, 
on the one hand because violations of Kolmogorov consistency conditions already manifest at the two-time level, which is therefore the minimal setting required to reveal multi-time nonclassicality.
On the other hand, the number of joint probabilities to be computed grows rapidly with the number of times, and multi-time experiments beyond two points become increasingly demanding, so that focusing on the two-time case provides the clearest and most tractable insight into the nonclassicality of quantum walks.
Hence, to quantify deviations from classical consistency conditions, we introduce the following Kolmogorov nonclassicality quantifier:
\begin{equation}\label{eq:kone}
    K(s,t) = \frac{1}{2}\sum_x \left| \, \sum_y P(x,t;y,s) - P(x,t) \, \right|,
\end{equation}
where $y$ is the outcome of the observable $X$ at the intermediate time $s$ 
and $x$ is the outcome of the same observable at the final time $t\geq s$. 
Now, $K(s,t)$ is the Kolmogorov distance \cite{Kolmogorov1963} between the probability distributions, respectively, $\{ \sum_y P(x,t;y,s)\}_x$ 
and $\{P(x,t)\}_x$.
Operationally, the former can be collected by performing both the measurements at time $s$ and at time $t$ and
then summing over the possible values of the intermediate measurement, while the latter
is associated with a single measurement at time $t$.
Furthermore, since $K(s,t)$ is a distance between probability distributions, it follows that
\begin{equation}
    0 \le K(s,t) \le 1 ,
\end{equation}
with $K(s,t)=0$ if and only if the two distributions coincide, i.e., if the measurement at time $s$ is effectively noninvasive at the level of outcome statistics at time $t$ and the Kolmogorov consistency condition is satisfied. 

Indeed, the two-time quantifier $K(s,t)$ depends on the choice of the intermediate time $s$.
To remove this arbitrariness and to ease the comparison with the one-time quantum-classical distance $D_{\mathrm{QC}}(t)$, we consider its time-averaged value over $s \in [0,t]$,
\begin{equation}\label{eq:kbard}
    \bar{K}(t) = \frac{1}{t} \int_{0}^{t} K(s,t)\, ds .
\end{equation}
Operationally, $\bar{K}(t)$ is the expected Kolmogorov violation when the intermediate measurement time is drawn uniformly at random in the interval $[0,t]$. Since $K(s,t)\ge 0$ for all $s\in[0,t]$, a strictly positive value $\bar{K}(t)>0$ certifies genuine two-time nonclassicality within $[0,t]$.

\subsection{Open-system multi-time probabilities and quantum regression formula}

When dealing with decoherence, i.e., when considering the walker as an open system,
we will restrict to the Markovian regime, where the dynamics is generated by the Lindblad master equation (\ref{eq:Lindblad}).
Not only that, but we will also express the multi-time probabilities using the quantum regression formula~\cite{Lax1968,Swain1981,Carmichael1993,Breuer2002,Guarnieri2014}, 
which ensures that the multi-time probabilities evolve under the same dynamical map that governs the quantum state evolution;
the validity of the quantum regression formula can be seen as a strong requirement of Markovianity \cite{Li2018}.

Explicitly, for projective measurements at times $s$ and $t\geq s$ of the observable $X$ of an open system
whose dynamics is fixed by the Lindblad generator $\mathcal{L}$, the quantum regression formula reads
\begin{equation}\label{eq:mainone}
    P(x,t;y,s)= \mathrm{Tr}\!\left[\mathcal{M}_x\, e^{\mathcal{L}(t-s)} \mathcal{M}_ye^{\mathcal{L}s}\rho(0)\right];
\end{equation}
note that, compared to the closed-system definition in Eq.(\ref{eq:multi}),
we simply replaced the unitary evolution with the superoperator $e^{\mathcal{L}t}$ that fixes the evolution of the open-system state --
a simple recipe that cannot be applied beyond the Markovian regime \cite{Guarnieri2014,Li2018}.
The one-time probability associated with a single measurement at time $t$ is indeed given by
\begin{equation}
    P(x,t) = \mathrm{Tr}\!\left[\Pi_x\, e^{\mathcal{L}t}\rho(0)\right].
\end{equation}

Introducing the full-dephasing superoperator $\Delta$, 
\begin{equation}\label{eq:fdeph}
    \Delta\rho = \sum_y \Pi_y \rho \Pi_y ,
\end{equation}
which describes the effect of a non-selective projective measurement of $X$ at time $s$,
and using the semigroup composition law
\begin{equation}
    e^{\mathcal{L}t} = e^{\mathcal{L}(t-s)}e^{\mathcal{L}s} \quad t\geq s\geq0,
\end{equation}
the two-time Kolmogorov quantifier in Eq.~\eqref{eq:kone} can be equivalently written as
\begin{equation}\label{eq:maintwo}
K(s,t)= \frac{1}{2}\sum_x \left|\mathrm{Tr}\!\left[\Pi_x\, e^{\mathcal{L}(t-s)} \big( \rho(s) -\Delta\rho(s)\big)\right] \right|.
\end{equation}
This equation is the starting point of the evaluation of $K(s,t)$, along with its time-averaged version $\bar{K}(t)$,
that will be carried out in the next sections
for CTQWs on different graphs,
both in the unitary case (where simply $\mathcal{L}\rho = - i [H,\rho]$) and in the presence of Markovian decoherence.

Before doing that, we note that Eq.~\eqref{eq:maintwo} implies that $K(s,t)$ is convex in its dependence on the initial state. 
Given $\rho(0) = \lambda \rho_1(0) + (1-\lambda)\rho_2(0)$ with $0\le\lambda\le1$, linearity of the dynamics and of the trace implies that the corresponding probabilities are convex combinations of those generated by $\rho_1(0)$ and $\rho_2(0)$ and then, since the Kolmogorov distance is convex with respect to each of its arguments, one has indeed
\begin{equation}\label{eq:convex}
    K_{\lambda\rho_1+(1-\lambda)\rho_2}(s,t)
    \leq \lambda K_{\rho_1}(s,t) + (1-\lambda)K_{\rho_2}(s,t).
\end{equation}
For this reason, we will always focus on the case of an initial site-localized state, 
$\rho(0)=\ket{\nu}\bra{\nu}$.
Besides ensuring $K(0,t)=0$ (since $[\Pi_x,\rho(0)]=0$), thus isolating temporal nonclassicality from preparation-induced one,
due to Eq.(\ref{eq:convex}) the choice of an initial localized pure states captures the extremal behavior of the nonclassicality quantifier.

\section{Multi-time nonclassicality in unitary quantum walks}\label{sec:kni}
We first analyse the nonclassicality of quantum walks in the case of a closed-system unitary dynamics. 
We study the Kolmogorov nonclassicality quantifier $K(s,t)$, along with its time average $\bar{K}(t)$, 
and we determine its topology-independent short-time behavior, as well as intermediate and long-time behaviors
that instead carry signatures of the specific topology of the walk, further comparing the multi-time nonclassicality 
with the single-time nonclassicality as quantified by the quantum-classical dynamical distance $D_{\mathrm{QC}}(t)$.

\subsection{Universal short-time behavior}
Expanding the unitary evolution operator to second order in time, we see that the one-time probabilities at short times are given by 
\begin{align}
    P(x,t) &= \left|\bra{x}e^{-i \lap t}\ket{\nu}\right|^2\label{eq:aux11}\\
    &= \delta_{x\nu} - \delta_{x\nu} (L^2)_{x\nu}t^2 + L^2_{x \nu} t^2 +\mathcal{O}(t^3),\nonumber
\end{align}
while the two-time probabilities can be written as
\begin{align}
   & P(x,t; y,s) = \left|\bra{x}e^{-i \lap (t-s)}\ket{y}\right|^2 \left|\bra{y}e^{-i \lap s}\ket{\nu}\right|^2 \nonumber\\
   & = \delta_{xy}\delta_{y\nu} - \delta_{y\nu} \left(\delta_{xy}(L^2)_{xy}-L^2_{xy}\right)(t-s)^2 \nonumber\\
  & -\delta_{xy}\left( \delta_{y\nu}(L^2)_{y\nu}-L^2_{y \nu}\right)s^2 +\mathcal{O}(t^3),
  \label{shorttime}
\end{align}
so that its marginal reads
\begin{align}
   \sum_y P(x,t; y,s) = & \delta_{x\nu} -\delta_{x \nu}  (L^2)_{x\nu}(t^2+2s^2-2 s t) \nonumber\\
   &+ L^2_{x\nu} (t^2+2s^2-2 s t) + \mathcal{O}(t^3).\label{eq:aux3}
\end{align}
Replacing Eqs.(\ref{eq:aux11}) and (\ref{eq:aux3}) in the definition of $K(s,t)$ -- see Eq.(\ref{eq:kone}) -- 
and using $(L^2)_{xx}=d_{x}+d_{x}^2 $ and 
$\sum_{x\neq \nu}L_{x\nu}^2 =d_{\nu}$,
we obtain the universal short-time behaviour of the nonclassicality quantifier:
\begin{equation}\label{eq:shorttimeK}
    K(s,t) = 2d_{\nu}\bigl(st - s^2\bigr) + \mathcal{O}(t^3),
\end{equation}
along with the short-time scaling of the time-averaged version of the quantifier of the violation of the Kolmogorov conditions -- see Eq.(\ref{eq:kbard}) --
\begin{equation}\label{eq:stq}
    \bar{K}(t ) =\frac{1}{3}d_\nu t^2 + \mathcal{O}(t^3).
\end{equation}
At very short times the walker has only started to delocalize from the initial site $\nu$. As a consequence, the unitary dynamics builds coherences between $\ket{\nu}$ and its nearest neighbours. The larger the number of such neighbours, i.e., the higher the local degree $d_\nu$, the richer this initial coherent superposition and the stronger the ensuing violation of Kolmogorov consistency when an intermediate measurement is performed. Remarkably, in this early-time regime the evolution is insensitive to global features of the graph: different topologies sharing the same local degree yield the same leading nonclassicality. Therefore, short-time Kolmogorov nonclassicality is topologically universal and entirely controlled by local connectivity. The parabolic dependence on $s$, peaking at $s=t/2$, reflects that the disturbance induced by the intermediate measurement is maximal when it occurs halfway through the evolution.

\subsection{Beyond short times: topology-driven behavior}
The previous analysis showed that, in the short-time regime, the degree of the initial node plays a key role in enhancing multi-time nonclassicality, as captured by the Kolmogorov-based quantifiers. 
This might suggest that increasing the overall connectivity or the number of nodes might generally enhance nonclassicality in quantum walks. 
However, this intuition does not hold at longer times. In fact, the behavior of the Kolmogorov-based quantifiers at larger temporal scales depends on the  topology of the graph.

\begin{figure}[t]   
  \centering
  \begin{subfigure}[b]{0.52\textwidth}
      \includegraphics[width=0.8\textwidth]{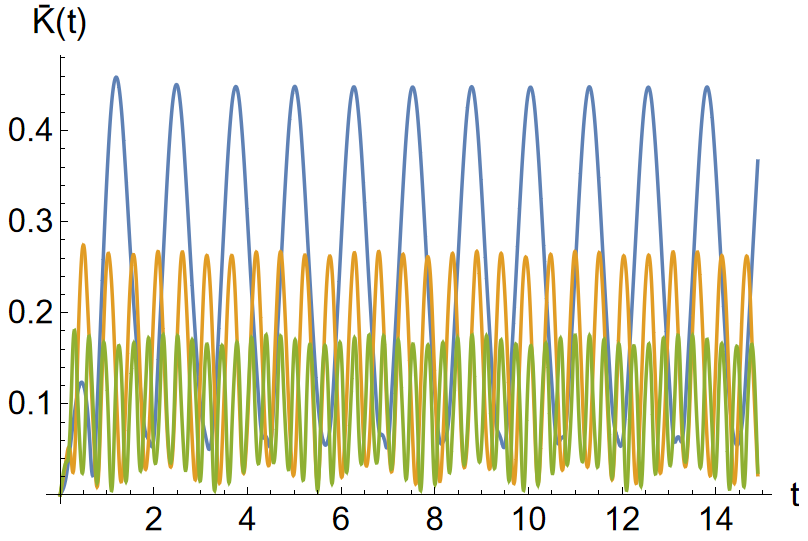}
    \caption{complete}
    \label{fig:KvsD complete}
  \end{subfigure}
  \begin{subfigure}[b]{0.46\textwidth}
      \includegraphics[width=0.85\textwidth]{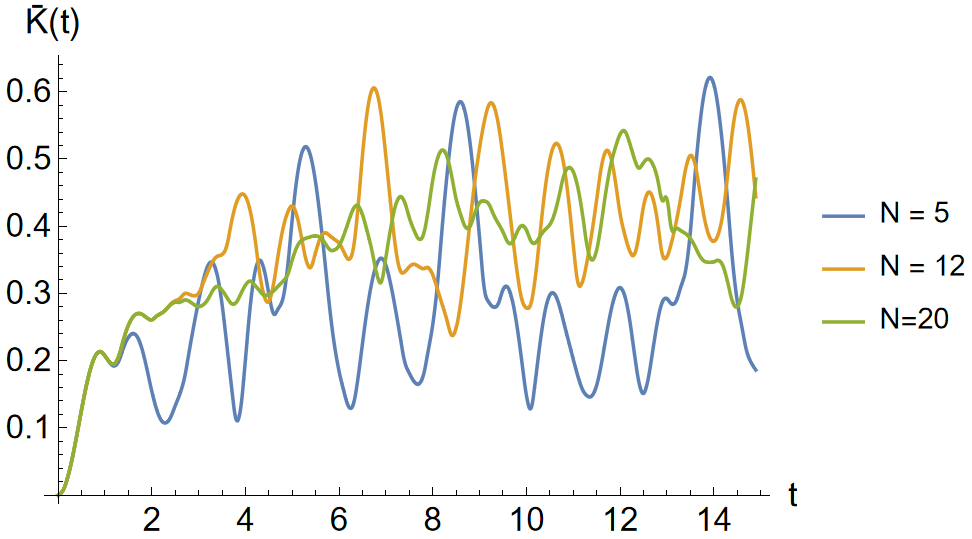}
    \caption{cycle}
    \label{fig:KvsDcycle}
  \end{subfigure}  
  \caption{Time-averaged Kolmogorov nonclassicality \( \bar{K}(t) \) as a function of the number of nodes \( N \), for complete graphs (a) and cycle graphs (b),
  in the case of a unitary dynamics.}
  \label{fig:KbarVsN}
\end{figure}

In particular, we study how the time-averaged quantity $\bar{K}(t)$  varies with the number of nodes $N$ for the two graph topologies introduced earlier, namely the complete  and the cycle graphs, 
see Fig.~\ref{fig:KbarVsN}.
In the complete graph, increasing the number of nodes leads to a clear reduction in the average nonclassicality. 
This behavior originates from the fact that, as $N$ grows, the walker tends to remain in the initial node with increasingly high probability, thereby suppressing temporal interference effects. 
This trend is consistent with the analytic expression of the transition probabilities for a CTQW on a complete graph \cite{mulken2011}:
\begin{align}
   \left|\bra{x}e^{-i L t} \ket{y}\right|^2 =
    \begin{cases} 
      \frac{4}{N^2}\sin^2\left(\frac{N}{2}t\right) \equiv P_{out}(t)  & \text{for }   x \neq y \\
     1 - (N-1)P_{out}(t) &\text{for }   x= y
   \end{cases}.
\end{align}
In fact, using these transition probabilities, the Kolmogorov quantifier can be written as
\begin{eqnarray}
    K(s,t)&=&(N-1)|P_{out}(t-s)+P_{out}(s)\nonumber\\
&&-N P_{out}(t-s)P_{out}(s) -P_{out}(t)|,
\end{eqnarray}
which leads to 
\begin{equation}\label{eq:aux7}
\begin{split}
    K(s,t)&=(N-1)\frac{16}{N^3}\left|
\frac{N}{4}\left[
\sin^2\left(\frac{N(t-s)}{2}\right)
+
\sin^2\left(\frac{Ns}{2}\right)\right.\right.\\
&\left.\left.-\sin^2\left(\frac{Nt}{2}\right)
\right]
- \sin^2\left(\frac{N}{2}(t-s)\right)\sin^2\left(\frac{N}{2}s\right)
\right|.
\end{split}    
\end{equation} 

In the large-$N$ limit, the oscillations of the transition probabilities $P_{out}(t)$ are suppressed, and the system tends to remain in the initial state with high probability. As a consequence,  classical features dominate, and the Kolmogorov nonclassicality,
along with its time average, approaches to zero as $1/N$ as $N$ increases.

Conversely, for the cycle graph, the time-averaged nonclassicality \( \bar{K}(t) \) exhibits persistent oscillations that do not vanish for large $N$, indicating a more robust form of nonclassical temporal correlations.
This  behavior can be explained by considering the explicit expression for the transition probabilities of the CTQW on the cycle graph. 
Using the expressions of eigenvalues and eigenvectors in Eqs. \eqref{eigenstatesCirc} and \eqref{lcycl} one finds \cite{mulken05}:
\begin{equation}
   \left|\bra{x}e^{-i L t} \ket{y}\right|^2 = \frac{1}{N^2}\left|
    \sum_{k=0}^{N-1} e^{-2i\left(1-\cos(\frac{2\pi k }{N})\right)t} e^{\frac{2i\pi k(x-y)}{N}} \right|^2,
\end{equation}
which converge to
\begin{equation}
    \lim_{N \to \infty}   \left|\bra{x}e^{-i L t} \ket{y}\right|^2 = \left|J_{x-y}(2 t) \right|^2,
\end{equation}
where $J_{n}(\tau) = \frac{1}{2\pi}\int_{-\pi}^{\pi}e^{i(\tau \sin \theta-n\theta)}d\theta$ is the Bessel function of the first kind.
This shows that their temporal oscillations persist for arbitrarily large $N$, so that nonclassical temporal correlations remain visible in the cycle graph.

 %The qualitative difference between the two topologies  reflects the fact that the complete graph quickly  suppresses the multi-time interference as the system size grows, whereas the cycles preserve quantum interference effects over longer timescales, even for larger $N$.
These results demonstrate that higher connectivity does not necessarily correspond to stronger temporal nonclassicality. Rather, it is the graph topology, 
not just the number of edges, that determines the persistence or suppression of temporal nonclassicality in the long-time regime.
Our results indeed point to the fact that nonclassicality of a quantum walk is not a unique notion.
Single-time and multi-time criteria probe different resources that can lead to different assessments of the same physical system.
The high connectivity of complete graphs is favorable in quantum algorithms, such as the spatial quantum search \cite{quantumsearch}, that involve single-time measurements. By contrast, if one is interested in the persistence of temporal quantum correlations, our results show that complete graphs suppress multi-time nonclassicality as captured by \(K(s,t)\) and its time average \(\bar K(t)\), whereas cycles sustain long-lived temporal nonclassicality. This suggests that cycles may be more favorable for sequential-measurement tasks that benefit from robust multi-time nonclassicality, such as randomness certification from temporal correlations \cite{Nath2024}.

\subsection{Comparison with the quantum–classical distance}
To assess in a quantitative way the difference between single-time and multi-time criteria of nonclassicality,
we now compare the time averaged Kolmogorov nonclassicality $\bar{K}(t)$ 
with the quantum–classical dynamical distance $D_{\mathrm{QC}}(t)$ recalled in Sec.\ref{sec:qcdd}.

First, we show that the two notions of nonclassicality do not satisfy a hierarchy, 
meaning that neither of the two implies the other.
Consider the simplest scenario of a two-site system with Laplacian 
\begin{equation}\label{eq:l2}
    L=\begin{pmatrix}1&-1\\-1&1\end{pmatrix},
\end{equation}
and localized initial state, say $\ket{\nu}=\ket{1}$.
The one-time probability for the CTQW is given by
\begin{equation}
   P(1,t) =  \left|\bra{1}e^{-i L t} \ket{1}\right|^2 = \tfrac12\bigl(1+\cos(2 t)\bigr), 
\end{equation}
and $P(0,t) = 1- P(1,t)$;
for the classical random walk on the same graph, the transition probabilities are instead given by -- see Eq.(\ref{eq:toc}) --
\begin{equation}
p_{1,1}(t) = \bra{1}e^{-L t} \ket{1} =\tfrac12\bigl(1+e^{-2t}\bigr)
\end{equation}
and $p_{0,1}(t) = 1- p_{1,1}(t)$.
At time $t^* (\approx 0.65)$ such that $e^{-2 t^*} = \cos(2t^*)$,
the quantum and classical distributions are then equal, so that the quantum-classical distance -- see Eq.(\ref{qcd}) --
is equal to zero, $D_{\mathrm{QC}}(t^\ast)=0$.
For the same model, we have the Kolmogorov quantifier
\begin{equation}\label{eq:kex}
    K(s,t)= \frac{1}{2}\left| \sin(2 s) \sin(2(t-s)) \right|,
\end{equation}
and its time average for $t \in [0, \pi/2]$ reads
\begin{equation}
\bar{K}(t) =\frac{1}{8 t}\Bigl(\sin(2 t)-2 t\cos(2 t)\Bigr),
\end{equation}
so that $\bar{K}(t^*) \neq 0$. 
We conclude that even though the quantum and classical random walks fixed by Eq.(\ref{eq:l2}) provide the same one-time probability distributions
at time $t^*$, there is no classical process whose (two-time) statistics reproduces the quantum statistics associated with sequential
measurements of the walker position, i.e., Eq.(\ref{eq:multi}), as testified by the violation of the Kolmogorov consistency conditions.

Conversely, $\bar{K}(t)=0$ certifies multi-time classicality,
i.e., existence of a classical stochastic model reproducing \emph{all} multi-time statistics.
However, $D_{\mathrm{QC}}(t)$ compares the CTQW with the \emph{fixed} random walk $e^{-Lt}$ on the same graph:
the classical model realizing $\bar{K}(t)=0$ needs not coincide with that particular random walk;
an explicit example of this is given in Sec.~\ref{subsec:haken-strobl}.

\begin{figure}[H]
    \centering
    \includegraphics[width=0.5\textwidth]{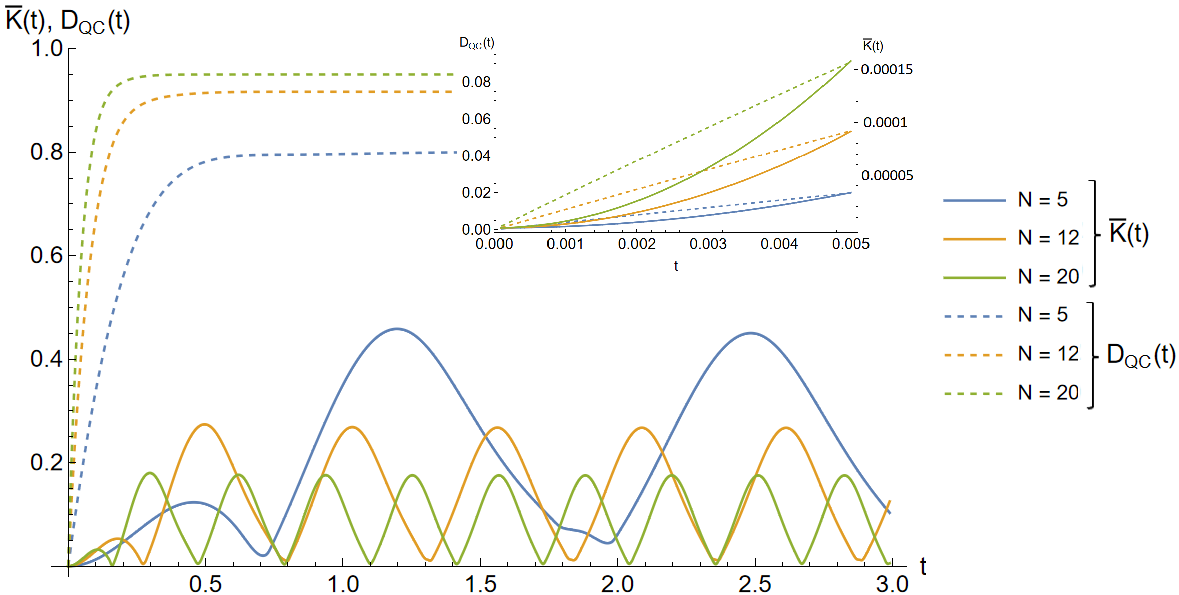}
    \caption{Time-averaged Kolmogorov nonclassicality $\bar{K}(t)$ (solid lines) and dynamical distance $D_{\mathrm{QC}}(t)$ (dashed lines) for complete graphs with different $N$. 
\textit{(Inset)} Short-time behavior.}
    \label{fig:KvsD}
\end{figure}

Going back to the general short-time behavior of the Kolmogorov quantifier, see Eq.(\ref{eq:stq}),
we note that it shares with the quantum-dynamical distance $D_{\mathrm{QC}}(t)$ its universal nature and dependence only on the degree of the initial node  $d_\nu$ in the short-time regime, but $D_{\mathrm{QC}}(t)$ scales linearly with time \cite{QCD}, $D_{\mathrm{QC}}(t) = d_\nu t + \mathcal{O}(t^2)$,
rather than quadratically.
On the other hand, their asymptotic behaviours are markedly different. For long times, the dynamical distance converges 
to a topology-independent value \cite{QCD}, 
$D_{\mathrm{QC}}(t) \longrightarrow 1 - \frac{1}{N}$ for $t\rightarrow \infty$, solely determined by the size of the system, 
while the behavior of $ \bar{K}(t) $ at long times is strongly influenced by the global topology of the graph. 
As previously discussed, for the complete graph increasing the number of nodes may reduce the multi-time nonclassicality, highlighting a nontrivial and topology-sensitive structure underlying temporal quantum correlations.
These findings are summarized in Fig.\ref{fig:KvsD}, where $D_{\mathrm{QC}}(t)$
and $\bar{K}(t)$ are plotted as a function of time for different values of $N$.

\section{Decoherent quantum walks}\label{sec:oqw}
We now extend our analysis to continuous-time quantum walks in an open system setting, where the dynamics is non-unitary due to interactions with an external environment.
This framework enables us to explore how decoherence and dissipation affect temporal nonclassicality, as captured by the Kolmogorov quantifiers.

\subsection{Decoherence in the position basis}
\label{subsec:haken-strobl}
We first consider dephasing in the position-basis, described by the Haken--Strobl model introduced in Sec.~\ref{sec:ctqw}, 
which provides a natural testbed to study how site-basis decoherence impacts temporal nonclassicality.

Starting with the short-time regime, we expand the density operator in powers of time,
\begin{equation}
    \rho(t) = e^{\mathcal{L}t}\rho(0) 
    = \rho(0) + t\,\mathcal{L}\rho(0) + \frac{t^2}{2}\mathcal{L}^2\rho(0) + \frac{t^3}{3!}\mathcal{L}^3\rho(0) + \dots
\end{equation}
For an initially localized state $\rho(0) = \ket{\nu}\bra{\nu}$, the first-order term reduces to the unitary contribution,
$-i[\lap,\rho(0)]$,
so that the dissipative part of the Lindblad generator starts contributing nontrivially at second order. In particular, one has
\begin{equation}
    \mathcal{L}^2\rho(0) = -[\lap,[\lap,\rho(0)]] - i\gamma\sum_{k\in \mathrm{NN}(\nu)}\left(\ket{k}\bra{\nu} - \ket{\nu}\bra{k}\right),
\end{equation}
where the sum runs over the nearest neighbors of the initial node $\nu$, $\mathrm{NN}(\nu)$.  
In fact, at short times coherences are generated only between the initial node and its directly connected neighbours, and it is precisely these newly formed coherences that are damped by the dissipator. Nevertheless, these corrections remain confined to the off-diagonal entries of $\rho(t)$ at this order, 
and therefore do not affect the measured probabilities. 
The first corrections to the populations and hence to the Kolmogorov quantifier appear only at third order, via the mixed pathway that transfers coherence loss into population dynamics.
Directly referring to the time-averaged quantifier, the short-time analysis -- see Appendix \ref{app:dit} -- provides us with the expansion
\begin{equation}\label{eq:kbtd}
    \bar{K} (t)
= \frac{1}{3}d_\nu t^2 - \frac{1}{6} \gamma d_\nu t^3 + O(t^4);
\end{equation}
as anticipated, decoherence affects violation of the Kolmogorov conditions from the third order in time, and it leads to a reduction of Kolmogorov nonclassicality. 

\begin{figure}[h!]
  \centering
  \begin{subfigure}[b]{0.51\textwidth}
    \includegraphics[width=\textwidth]{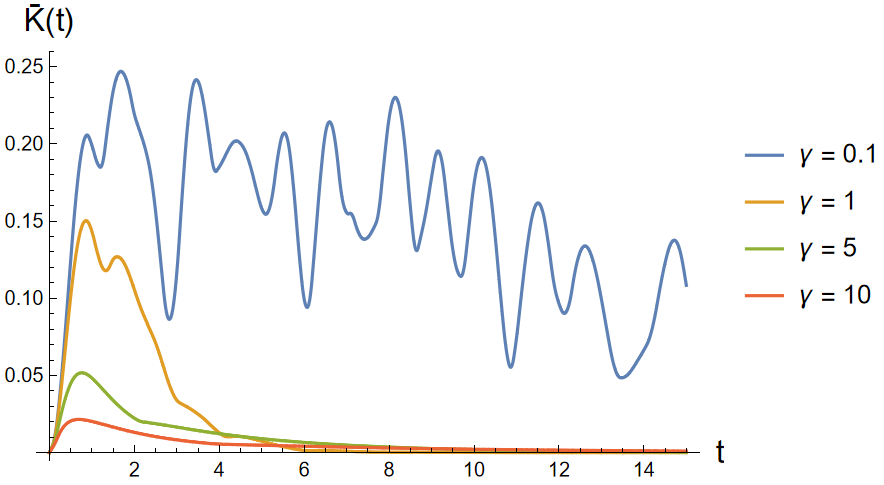}
    \caption{cycle}
    \label{fig:}
  \end{subfigure}
  \begin{subfigure}[b]{0.43\textwidth}
    \includegraphics[width=\textwidth]{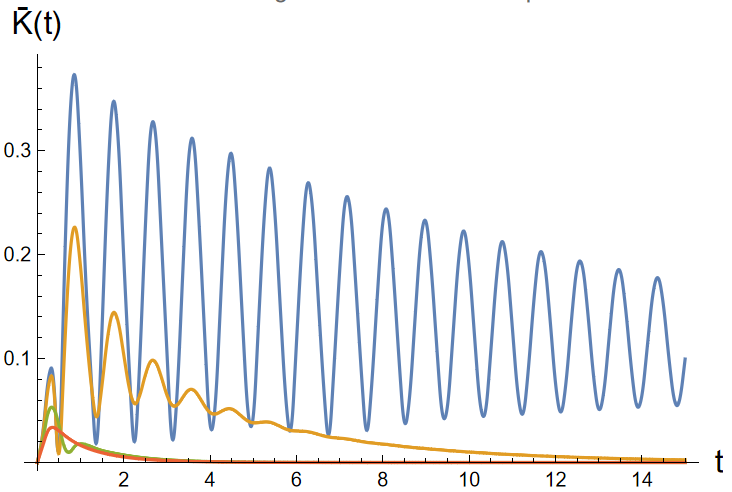}
    \caption{complete}
    \label{fig:completeNt15}
  \end{subfigure}
  \caption{Kolmogorov nonclassicality under Haken--Strobl dephasing for cycle (a) and complete (b) graphs,
  for $N = 7$.}
  \label{fig:Haken}
\end{figure}

Moving beyond the short-time regime, the complete behavior of the Kolmogorov nonclassicality quantifier $\bar{K}(t)$ is shown in Fig.\ref{fig:Haken}, 
for the cycle and complete graph. We can observe the damping induced by decoherence -- compare with Fig.\ref{fig:KbarVsN} --
and that the differences between the two topologies are progressively less pronounced for stronger dephasing rates. 
Importantly, in both cases we have a convergence to a classical behavior on long times.
In fact, the time-averaged Kolmogorov quantifier satisfies the bound, see Appendix \ref{app:dit},
\begin{equation}
    \bar K(t)\le \frac{\sqrt{N}}{\mu_2 t}\left(1- e^{-\mu_2 t} \right),
\end{equation}
where $-\mu_2<0$ is the spectral gap of the Lindbladian $\mathcal{L}$ -- i.e., the largest nonzero real part of the eigenvalues of $\mathcal{L}$  -- with, 
see Eqs.(\ref{eq:Lindblad}) and~(\ref{eq:Haken}),
\begin{equation}
\mathcal{L}[\rho] = -i[L, \rho] + \gamma(\Delta - \mathbb{I})[\rho].
\end{equation}
This means in particular that position-basis decoherence suppresses the multi-time nonclassicality identified by the Kolmogorov consistency conditions
for any graph topology, on a time scale that is at most $1/\mu_2$.
While in the weak-dephasing regime $\mu_2$ is expected to be of the order of $\gamma$, 
for strong depashing, coherences decay on a much shorter time scale and the dynamics becomes effectively classical at the level of populations, 
so that $\mu_2$ depends on the interplay between the Laplacian spectral gap and the dephasing rate \cite{catalano23}. 
More specifically, for $\gamma \gg \lambda_F$ -- where $\lambda_F$ is the Fiedler value (second smallest eigenvalue of $L$) --
adiabatic elimination yields an effective classical diffusion for the populations, $\dot{{\bf p}}(t) = -(2/\gamma)L {\bf p}(t)$, whose slowest relaxation mode scales as 
$\mu_2 \sim 2\lambda_F / \gamma$.

While the long-time convergence to classicality for position decoherence is shown also by the quantum-classical distance \cite{DQCdecoherence},
the transient behaviors of $\bar{K}(t)$ and $D_{\mathrm{QC}}(t)$ can be radically different.
To illustrate this, let us consider the limit situation where decoherence acts so strongly that it instantly destroys 
coherences - so that the open-system evolution coincides with the full dephasing map $\Delta$.
Indeed, any initial state that is diagonal in the position basis will remain diagonal -- actually it will not evolve at all --
neither it will be modified by a nonselective measurement of the position, so that $\overline{K}(t) = 0$ for any $t$.
On the other hand, $D_{\mathrm{QC}}(t)$ is non-monotonic as a function of $\gamma$: it decreases with $\gamma$ up to a threshold value and then increases again; even when decoherence dominates over the unitary dynamics and effectively freezes the walker near its initial node, the classical random walk continues to spread over the graph, until it reaches the steady-state distribution.

\subsection{Decoherence in the energy basis}

As second case study of the impact of decoherence, we consider pure dephasing in the energy eigenbasis of the Hamiltonian;
crucially, in contrast with the previous case, now decoherence occurs on a basis that is different from the measurement basis.

Again, we start from the short-time regime and at first order in time we have 
\begin{equation}
    \mathcal{L}\rho(0)=-i[\lap,\rho(0)]-\frac{\gamma}{2}[\lap,[\lap,\rho(0)]] + O(t^2).
\end{equation}
In this case, decoherence already affects the probabilities at first order in time
and, as shown in the Appendix \ref{app:dite}, the Kolmogorov quantifiers are influenced by decoherence already at second order,
%Therefore, even though the probabilities are affected by dephasing, the Kolmogorov quantifier may deviate from the unitary case at the same order, or---as in this case---at a higher order in time compared to the probabilities. Decoherence effects become visible in $K(s,t)$ already at second order, 
in contrast with the position-basis case where they appear only at third order. With site dephasing, the dissipator must first wait for the unitary dynamics to create coherences, then damp them, and finally feed changes back into the populations -- hence the extra power of $t$. 
With energy dephasing, by contrast, decoherence acts directly on the energy coherences to which $K$ is sensitive; symmetry removes the linear term, leaving an $O(t^2)$ correction.
Explicitly, the short-time expansion reads 
\begin{equation}
\bar{K}(t)= \frac{1}{12}\sum_x \left |A_x + \frac{\gamma^2}{4}B_x \right| t^2
+ O(t^3),
\end{equation}
where we introduced the quantities
\begin{eqnarray}
A_x &=& \bra{x}\mathcal{L}_H^2[\rho(0)]\ket{x}, \nonumber\\
B_x &=& 
\bra{x}\mathcal{L}_H^2\!(\mathbb{I}-\Delta)\,\mathcal{L}_H^2[\rho(0)]\ket{x},
\end{eqnarray}
with $\mathcal{L}_H [\rho]= -i [L, \rho]$ the Hamiltonian part of the dissipator;
note that %in this case the lowest non-trivial order in $t$ already shows the effect of the dissipative term in the Lindblad generator.
the sum $\sum_x |A_x|$ is exactly $4d_{\nu}$, then for $\gamma=0$ we recover the unitary expansion of $\bar{K}(t)$.

\begin{figure}[h!]
  \centering
  \begin{subfigure}[b]{0.53\textwidth}
    \includegraphics[width=0.9\textwidth]{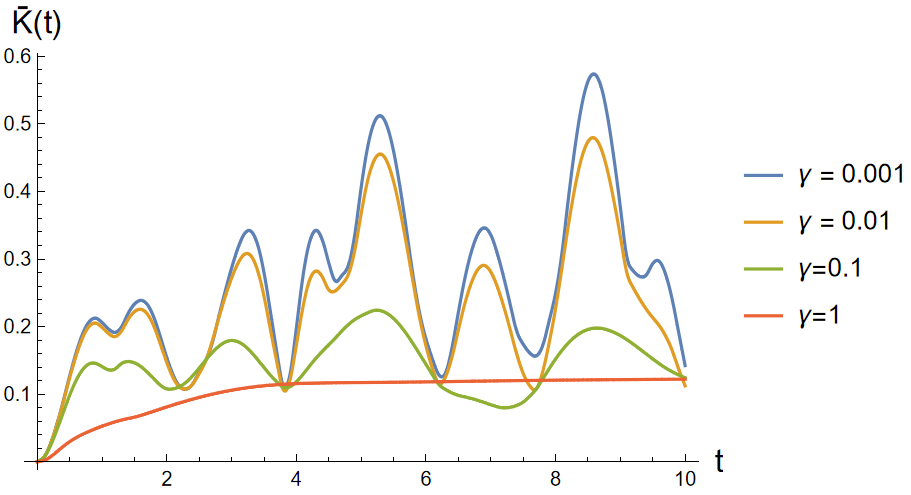}
    \caption{cycle}
    \label{fig:}
  \end{subfigure}
  \begin{subfigure}[b]{0.41\textwidth}
    \includegraphics[width=\textwidth]{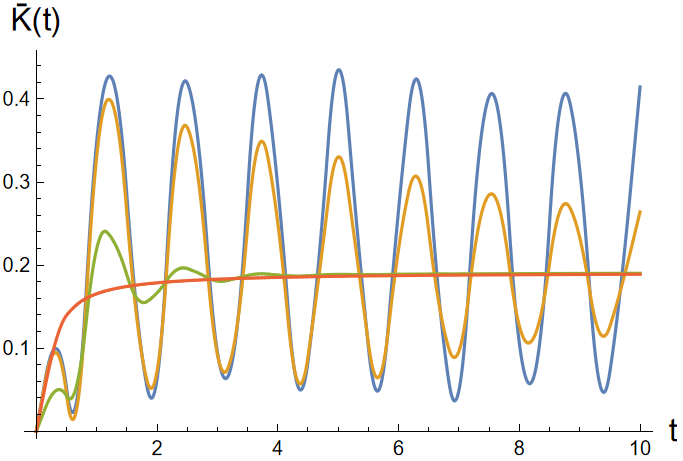}
    \caption{complete}
    \label{fig:}
  \end{subfigure}
  \caption{Kolmogorov nonclassicality under decoherence in the energy basis for cycle (a) and complete (b) graphs with $N=5$ nodes.}
  \label{fig:decoerenzaenergia}
\end{figure}

The full behavior of the Kolmogorov nonclassicality quantifier is shown in Fig.\ref{fig:decoerenzaenergia}. While the oscillations
are suppressed more and more with growing value of the dephasing rate $\gamma$, nonzero nonclassicality can survive on long times,
for both the cycle and the complete graphs.
In Appendix \ref{app:dite}, we derive closed-form expressions for the asymptotic values of the time-averaged Kolmogorov quantifier. For a complete graph, we get
\begin{equation}\label{eq:kbinf}
\lim_{t\to\infty}\bar K(t)=\frac{2\,(N-1)(N-2)}{N^3},
\end{equation}
while for a cycle graph
\begin{equation}\label{eq:asnn}
\lim_{t\to\infty}\bar K(t)=
\begin{cases}
\displaystyle \frac{\,(N-1)^2}{N^3}, & \text{$N$ odd},\\[6pt]
\displaystyle \frac{2\,(N-2)^2}{N^3}, & \text{$N$ even}.
\end{cases}
\end{equation}
In both cases, the asymptotic nonclassicality decreases with $N$ while remaining strictly positive for any finite value of $N>2$.

The nonzero asymptotic nonclassicality can be understood starting from Eq.(\ref{eq:soluzioneesattaenergy}): taking the limit
$\rho(\infty) = \lim\limits_{t \to \infty} \rho(t)$, all terms with $\lambda \neq \lambda'$ are exponentially suppressed and one obtains
\begin{equation}\label{eq:ltl}
\rho(\infty)=\sum_{\lambda}\Xi_{\lambda}\rho(0)\Xi_{\lambda},
\end{equation}
where the $\Xi_{\lambda}$ are the projectors into the Laplacian eigenspaces defined in Eq.(\ref{eq:xil}).
In other terms, intrinsic decoherence acts on long times as a full dephasing -- compare with Eq.(\ref{eq:fdeph}) -- 
but with respect to the Laplacian eigenbasis: coherences between different eigenspaces are erased, while coherences within degenerate eigenspaces are preserved. 
Importantly, even for a nondegenerate spectrum, the asymptotic state $\rho(\infty)$ will not be diagonal 
in the site basis -- that is, the measurement basis -- since Laplacian eigenvectors are typically delocalized. 
The site-basis coherences that survive at long times are the reason behind the violation of the Kolmogorov consistency conditions,
since their destruction by an intermediate (nonselective) measurement will have an impact on the subsequent statistics.
An explicit example of a nondegenerate Laplacian with residual nonclassicality $\bar{K}(t)$ for $t\longrightarrow \infty$
is reported in Appendix \ref{app:dite}.

Finally, note that for both the complete and the cycle graph the asymptotic value 
of $D_{\mathrm{QC}}(t)$ depends only on $N$, and it coincides with the asymptotic value in the unitary case for $N \to \infty$ \cite{DQCdecoherence}. 
Our results for $\bar{K}(t)$ display a similar feature: energy-basis dephasing suppresses the oscillatory contributions but leaves the $1/N$ scaling unchanged -- compare with Eq.(\ref{eq:aux7}). Thus, intrinsic decoherence primarily damps temporal oscillations without modifying the large-$N$ structure of either quantifier, whereas site dephasing drives both $D_{\mathrm{QC}}(t)$ and $\bar{K}(t)$ to zero. In particular, for intrinsic decoherence the contrast between the $\mathcal{O}(1)$ asymptotics of $D_{\mathrm{QC}}(t)$ and the vanishing $\mathcal{O}(1/N)$ behaviour of $\bar{K}(t)$ highlights the greater sensitivity of multi-time nonclassicality to noise and Hilbert-space dimension.
Notably, however, the
origin of the asymptotic values differs for the two quantifiers:
for intrinsic decoherence, a nonzero long-time $D_{\mathrm{QC}}$ is tied to
residual coherence within degenerate Laplacian eigenspaces (and thus disappears
in the absence of such degeneracies), whereas $\bar K(t)$ is
controlled by the overlap structure of Laplacian eigenspaces with the site
basis and can remain nonzero even for nondegenerate spectra.
%\\Altogether, the open-system analysis shows that different decoherence channels affect temporal nonclassicality in sharply different ways: site dephasing enforces Kolmogorov consistency at long times, while intrinsic dephasing can protect a finite $\overline{K}(\infty)$ whenever the Laplacian has degeneracies.

\section{Conclusions}\label{sec:cao}
In this work, we have investigated temporal nonclassicality in continuous-time quantum walks by means of a multi-time quantifier of the violations of the Kolmogorov consistency conditions, comparing it with a single-time quantifier of the distance between the dynamical evolutions of classical and quantum walks.
Our comparison underscores the operational nature of nonclassicality: its detection and quantification depend on the probing protocol, and single-time benchmarks versus multi-time tests may lead to markedly different assessments.

For unitary quantum walks, we showed that the Kolmogorov quantifier exhibits a short-time scaling that depends only on the local connectivity of the initial site, while its behavior at longer times is strongly influenced by the graph topology. In particular, we found qualitatively different regimes for cycle and complete graphs, reflecting the distinct spectral properties of their Laplacians. 
We then extended the analysis to open-system dynamics, focusing on two paradigmatic decoherence mechanisms. For site-basis dephasing described by the Haken–Strobl master equation, we derived a bound showing that the time-averaged Kolmogorov quantifier vanishes asymptotically for any connected graph. This confirms that, in this case, decoherence suppresses temporal nonclassicality in the long-time limit. In contrast, for energy-basis dephasing (intrinsic decoherence), we obtained closed-form expressions for the asymptotic value of the quantifier on both cycle and complete graphs, showing that a finite long-time value can persist.
The origin of this residual nonclassicality can be traced back to the structure of the Laplacian eigenvectors. Even when intrinsic decoherence eliminates coherences between different energy eigenspaces, it leaves the asymptotic state non-diagonal in the site basis since the eigenvectors are delocalized. 
These results clarify how different forms of decoherence affect multi-time quantum correlations in continuous-time quantum walks, and they highlight the sensitivity of temporal nonclassicality to both the graph topology and the nature of the environment-induced noise.

Several interesting directions for future investigation remain open. On the quantum walk side, it would be interesting to consider chiral quantum walks, i.e., quantum walks with complex phases in the Laplacian. This is particularly relevant for the study of multi-time statistics, as chirality breaks the time symmetry of transition probabilities and may lead to qualitatively new temporal signatures.
Another natural extension concerns the analysis of more general evolutions, in particular non-Markovian processes. 
The inclusion of memory effects in the multi-time statistics would require a more general approach, beyond the quantum regression formula,
but would allow for a more complete characterization of the complex interplay between temporal correlations and quantum behaviors.

\acknowledgements
This work has been supported by the Italian Ministry of Research and Next Generation
EU via the PRIN 2022 project Quantum Reservoir Computing (QuReCo)
(contract n. 2022FEXLYB), the PRIN 2022 Project EQWALITY (Contract N. 202224BTFZ),
and the NQSTI-Spoke1-BaC project QSynKrono (contract n. PE00000023-QuSynKrono).

\clearpage
\appendix
\onecolumngrid

\renewcommand{\theequation}{\thesection\arabic{equation}}
\setcounter{equation}{0}

\section{Decoherence in the position basis}\label{app:dit}

\subsection{Short-time regime}
We first derive the short-time expression of $K(s,t)$ in the presence of site dephasing governed by the Haken--Strobl master equation,
\begin{equation}
    \mathcal{L}[\rho(t)]=\mathcal{L}_H[\rho(t)] + \mathcal{D}[\rho(t)],
\end{equation}
where 
\begin{equation}
\mathcal{L}_H[\rho] = -i\,[\lap,\rho]
\end{equation}
is the unitary Liouvillian generated by a real symmetric Laplacian $L$, and
\begin{equation}
    \mathcal{D}[\rho] = \gamma(\Delta - \mathbb{I})[\rho]
\end{equation}
is the dissipator part of the generator, with
\begin{equation}
        \Delta[X]=\sum_k |k\rangle\!\langle k|\,X\,|k\rangle\!\langle k|
\end{equation}
the full dephasing map in the position basis. For any operator $X$ diagonal in the position basis, $\mathcal{D}[X]=0$ 
and $\mathcal{L}_H[X]$ is purely off-diagonal,
i.e., $\bra{x}\mathcal{L}_H[X]\ket{x} = 0$ for any $x$; for any off-diagonal $X$, $\mathcal{D}[X]=-\gamma X$.
Consider the localized initial state $\rho(0)=|\nu\rangle\!\langle\nu|$, which is diagonal in the position basis. Keeping track of whether terms are diagonal (Diag) or off-diagonal (OffDiag) in the measurement basis, i.e. position basis, one has:
\begin{equation}
    \begin{split}
        &\mathcal{L}[\rho(0)] = \mathcal{L}_{H}[\rho(0)] \quad\rm{(OffDiag)},\\
        &\mathcal{L}^2[\rho(0)]=\underbrace{\mathcal{L}^2_{H}[\rho(0)]}_{\rm{Diag+OffDiag}}+\underbrace{\mathcal{D}\mathcal{L}_{H}[\rho(0)]}_{-\gamma\mathcal{L}_{H}[\rho(0)]\rm{(OffDiag)}},\\
        &\mathcal{L}^3[\rho(0)]=\underbrace{\mathcal{L}^3_{H}[\rho(0)]}_{\rm{OffDiag}}-\underbrace{\gamma \mathcal{L}_{H}^2[\rho(0)]}_{\rm{Diag+OffDiag}}+\underbrace{\mathcal{D}\mathcal{L}^2_{H}[\rho(0)]}_{\rm{OffDiag}}+\underbrace{\mathcal{D}^2\mathcal{L}_{H}[\rho(0)]}_{\rm{OffDiag}}.
    \end{split}
\end{equation}
We used the fact that, for an initial state diagonal in the position basis, a real and symmetric Laplacian $L$, and measurements performed in the same position basis, all odd powers of the unitary Liouvillian produce purely off-diagonal operators,
\begin{equation}
    \mathcal{L}_H^{2m+1}[\rho(0)] \in \mathrm{OffDiag}, 
    \qquad \forall\, m \ge 0.
\end{equation}
To see this, note that $\mathcal{L}_H[\rho(0)] = -i\,\mathcal{C}_H[\rho(0)]$, where $\mathcal{C}_\lap[X] = [\lap,X]$ denotes the commutator with the Hamiltonian, and $\mathcal{C}_\lap^n[X]$ its $n$-fold application. 
Since both $\lap$ and $\rho(0)$ are real matrices, $\mathcal{C}_\lap^n[\rho(0)]$ is real for all $n$, so that
\begin{equation}
    \mathcal{L}_H^{n}[\rho(0)] = (-i)^n\,\mathcal{C}_H^n[\rho(0)]
\end{equation}
is purely imaginary for odd $n$. 
But then, since $\bra{x}e^{\mathcal{L}_H t}[\rho(0)]\ket{x}=\sum_{n}\frac{1}{n!}\bra{x}(\mathcal{L}_H t)^n[\rho(0)]\ket{x}$
must be real for all times, the contributions from the odd orders vanish. \\
Expanding $e^{\mathcal{L}\tau}$, with $\tau=t-s$, and $\rho(s)$ up to overall third order in time, the operatorial part in Eq.(\ref{eq:maintwo}) becomes
\begin{equation}
\begin{aligned}
    e^{\mathcal{L}\tau} \!\left[\rho(s)-\Delta[\rho(s)]\right]
    &= \Big(\mathbb{I}+\tau\mathcal{L}+\tfrac{\tau^2}{2}\mathcal{L}^2+\tfrac{\tau^3}{3!}\mathcal{L}^3\Big)
       \Big[\rho(0)+s\mathcal{L}[\rho(0)]+\tfrac{s^2}{2}\mathcal{L}^2[\rho(0)]+\tfrac{s^3}{3!}\mathcal{L}^3[\rho(0)] \\
    &\hspace{8.1em}
       -\,\Delta\!\left[\rho(0)+s\mathcal{L}[\rho(0)]+\tfrac{s^2}{2}\mathcal{L}^2[\rho(0)]+\tfrac{s^3}{3!}\mathcal{L}^3[\rho(0)]\right]\Big]
       \;+\; \mathcal{O}(t^4).
\end{aligned}
\end{equation}
%\end{equation}
Using $\rho(0)-\Delta[\rho(0)]=0, \operatorname{Tr}\!\left[|x\rangle\!\langle x|\,(Z-\Delta[Z])\right]=0$ for any operator $Z$,
and that $\mathcal{L}\Delta[Z]$ is off-diagonal for any $Z$, one obtains up to $\mathcal{O}(t^3)$ provided that $||\mathcal{L}t|| \ll 1$ \cite{Szigeti2019}:
\begin{equation}
    \bra{x}e^{\mathcal{L}\tau} \!\left[\rho(s)-\Delta[\rho(s)]\right]\ket{x}
    \;=\; \Big(\tau s - \tfrac{\gamma}{2}\,s\,\tau\,(\tau+s)\Big)\,\bra{x}\mathcal{L}_H^2[\rho(0)]\ket{x} \;+\; \mathcal{O}(t^4),
\end{equation}
so that, see Eq.(\ref{eq:maintwo}),
\begin{equation}
    K(s,t)
    \;=\; \frac{1}{2}(t-s) s\Big(1 - \tfrac{\gamma}{2}\,t\Big)\,
           \sum_{x}\left|\langle x|\,\mathcal{L}_H^2\!\big[|\nu\rangle\!\langle\nu|\big]\,|x\rangle\right|
           \;+\; \mathcal{O}(t^4),\label{eq:kst33}
\end{equation}
with
\begin{align}
\langle x|\mathcal{L}^2_H[\ketbra{\nu}{\nu}]|x\rangle
=-2(L^2)_{x\nu}\delta_{x\nu}+2L_{x\nu}^2.
\end{align}
Recalling that $(L^2)_{\nu\nu}=d_{\nu}+d_{\nu}^2 $ and that 
$\sum_{x\neq \nu}L_{x\nu}^2 =d_{\nu}$ , we can write:
\begin{align}\label{eq:sumdu}
\sum_x\left|\langle x|\mathcal{L}^2_H[\ketbra{\nu}{\nu}]|x\rangle\right|
=
4 d_{\nu}.
\end{align}
Hence, to third order in time,
\begin{align}\label{eq:ktt}
   & K(s,t) \;=\; 2d_\nu\,s\,(t-s) \;-\; \gamma\,d_\nu\,s\,(t-s)\,t \;+\; \mathcal{O}(t^4);
\end{align}
indeed, integrating this expression one gets Eq.(\ref{eq:kbtd}).
In the unitary limit ($\gamma=0$), Eq.(\ref{eq:ktt}) reduces to Eq.(\ref{eq:shorttimeK}).

\subsection{Asymptotic regime}

We now show that for any semigroup dynamics under the Haken-Strobl master equation (\ref{eq:Haken}), with any connected graph $L$, 
the time-averaged Kolmogorov quantifier $\bar{K}(t)$ is bounded by
\begin{equation}\label{eq:kbarp}
    \bar K(t)\le \frac{\sqrt{N}}{\mu_2 t}\left(1- e^{-\mu_2 t} \right),
\end{equation}
where $-\mu_2 < 0$ is the spectral gap of the Lindbladian $\mathcal{L}$, i.e., the largest nonzero real part of its eigenvalues.
Indeed, this guarantees in particular that there is no asymptotic Kolmogorov nonclassicality, $\bar K(t) \rightarrow 0$ for $t \rightarrow \infty$, for decoherence in the position basis.

\begin{proof}[Proof]
Starting from $K(s,t)$ as in Eq.(\ref{eq:maintwo}), we have 
\begin{equation}
\begin{split}
    K(s,t) 
    & =\frac{1}{2}\sum_x\left|\bra{x}\left(e^{\mathcal{L}(t-s)}[\rho(s)]-e^{\mathcal{L}(t-s)}\Delta[\rho(s)]\right)\ket{x}\right| 
    \le \frac{1}{2}\Big\|\ e^{\mathcal{L}(t-s)}[\rho(s)]-e^{\mathcal{L}(t-s)}[\Delta[\rho(s)]]\Big\|_{1} \\
    & \leq \frac{1}{2}\Big\|\  e^{\mathcal{L}(t-s)}[\rho(s)] - \frac{\mathbb{I}}{N}\Big\|_{1} + 
    \frac{1}{2}\Big\|\  e^{\mathcal{L}(t-s)}\Delta[\rho(s)] - \frac{\mathbb{I}}{N}\Big\|_{1}, 
    \end{split}
\end{equation}
where the first inequality, $\sum_x |\bra{x}A\ket{x}| \leq \|A\|_1$, follows from \cite{Nielsen2010} $|\mbox{Tr}\left[U A\right]| \leq \|A\|_1$
for all unitaries $U$,
and in the second we used the triangular inequality after summing and subtracting the fully mixed state
$\mathbb{I}/N$. 
The latter is the unique steady state for the generator defined by Eq.(\ref{eq:Haken}): 
Haken--Strobl dephasing forces any stationary state to be diagonal, 
while connectivity of the graph ensures uniqueness of $\mathbb{I}/N$.
In other terms, the resulting semigroup dynamics is primitive and we can then use the bound (see for example equation (1.2) of \cite{Bardet2022}):
\begin{equation}
\left\|e^{\mathcal{L} t}\left[\rho-\frac{\mathbb{I}}{N}\right]\right\|_1 \leq \sqrt{N}  e^{-\mu_2 t},
\end{equation}
where $-\mu_2 <0$ is the largest nonzero eigenvalue of the Lindbladian $\mathcal{L}$.
We thus end up with
\begin{equation}
     K(s,t) \leq \sqrt{N}e^{-\mu_2 (t-s)},
\end{equation}
and,
taking the average over the time $s$ -- see Eq.(\ref{eq:kbard}) -- we get Eq.(\ref{eq:kbarp}).
\end{proof}

\section{Decoherence in the energy basis}\label{app:dite}
\subsection{Short-time regime}
We derive the short-time expression of $K(s,t)$ in the presence of decoherence in the energy basis, described by the master equation:
\begin{equation}
\frac{d\rho(t)}{dt} =-i[\lap,\rho(t)] - \frac{\gamma}{2}[\lap,[\lap,\rho(t)]]
\end{equation}
where  $\gamma$ denotes the decoherence rate.
We start from the initial state $\rho(0) = \ketbra{\nu}{\nu}$ and we keep track
of the diagonal and off-diagonal terms in the short-time expansion of
$e^{\mathcal{L}t}\rho(0)$:
\begin{equation}
    \begin{split}
        &\mathcal{L}\rho(0) = \underbrace{\mathcal{L}_{H}\rho(0)}_{\rm{OffDiag}} +\frac{\gamma}{2}\underbrace{\mathcal{L}^2_{H}\rho(0)}_{\rm{Diag+OffDiag}} \\
        &\mathcal{L}^2\rho(0)=\underbrace{\mathcal{L}^2_{H}\rho(0)}_{\rm{Diag+OffDiag}}+\gamma\underbrace{\mathcal{L}^3_{H}\rho(0)}_{\rm{OffDiag}}+\frac{\gamma^2}{4}\underbrace{\mathcal{L}^4_{H}\rho(0)}_{\rm{Diag+OffDiag}}.\\
        &
    \end{split}
\end{equation}
Proceeding in a similar way as for decoherence in the position basis, we obtain:
\begin{equation}
K(s,t)= \frac{1}{2}s(t-s)\sum_x \left |A_x + \frac{\gamma^2}{4}B_x \right|
+ O(t^3)\; .
\end{equation}
where we introduced the quantities
\begin{equation}
A_x = \bra{x}\mathcal{L}_H^2[\rho(0)]\ket{x}, 
\qquad 
B_x =\bra{x}\mathcal{L}_H^2\!(\mathbb{I}-\Delta)\,\mathcal{L}_H^2[\rho(0)]\ket{x}.
\end{equation}

The sum $\sum_x |A_x|$ is exactly $4d_{\nu}$, see Eq.(\ref{eq:sumdu}), so that for $\gamma=0$ we recover the unitary expansion of $K(s,t)$;
moreover, note that in this case the lowest non-trivial order in $t$ already shows the effect of the dissipative term in the Lindblad generator.

\subsection{Asymptotic regime}\label{ap:asymptotic}

 Replacing Eq.(\ref{eq:soluzioneesattaenergy}) in Eq.(\ref{eq:maintwo}), we have
\begin{equation}
    K(s, t)=\frac{1}{2}\sum_x \left|\sum_{\lambda,\lambda',\lambda'',\lambda'''}
    e^{-i(\lambda-\lambda')s-\frac{\gamma}{2}(\lambda-\lambda')^2 s}e^{-i(\lambda''-\lambda''')(t-s)-\frac{\gamma}{2}(\lambda''-\lambda''')^2 (t-s)}
    \bra{x}\Xi_{\lambda''} \big( \mathbb{I} -\Delta\big)\left[\Xi_{\lambda}\rho(0)\Xi_{\lambda'}\right]\Xi_{\lambda'''}\ket{x} \right|
\end{equation}
and then taking the average, $\lim_{t \rightarrow \infty} \frac{1}{t}\int_0^t$, only the terms with $\lambda=\lambda'$ and $\lambda''=\lambda'''$
survive, i.e.,
\begin{equation}
    \lim_{t \rightarrow \infty} \bar{K}(t)= \lim_{t \rightarrow \infty} \frac{1}{t}\int_0^t d s K(s,t) = \frac{1}{2}\sum_x
    \left|\sum_{\lambda}\left|\bra{x} \Xi_{\lambda} \ket{\nu}\right|^2 
    -\sum_y\sum_{\lambda}\sum_{\lambda'}\left|\bra{x} \Xi_{\lambda'} \ket{y}\right|^2 
    \left|\bra{y} \Xi_{\lambda} \ket{\nu}\right|^2 
    \right|;
\end{equation}
it is then convenient to define the matrix
\begin{equation}\label{eq:aux0}
F_{x,y} = \sum_{\lambda}\left|\bra{x} \Xi_{\lambda} \ket{y}\right|^2,
\end{equation}
so that
\begin{equation}\label{eq:aux1}
    \lim_{t \rightarrow \infty} \bar{K}(t) = \frac{1}{2}\sum_x \left|(F- F^2)_{x,\nu} \right|.
\end{equation}

\subsection*{Complete graph}

For the complete graph, we have only two eigenvalues $\lambda=0$ and $\lambda=N$, with the corresponding projectors satisfying
\begin{equation}
\bra{x} \Xi_{0} \ket{y} = \frac{1}{N},
\quad \bra{x} \Xi_{N} \ket{y}  = \delta_{xy} - \frac{1}{N}.
\end{equation}
Hence,
\begin{equation}
F_{x,y} =
\begin{cases}
\displaystyle \dfrac{2}{N^2}- \dfrac{2}{N} + 1, & x = y,\\[6pt]
\displaystyle \dfrac{2}{N^2}, & x\neq y,
\end{cases}
\end{equation}
i.e., the matrix $F$ can be written in terms of the all-ones matrix $J$ and the identity matrix $\mathbb{I}$ as
\begin{equation}
     F = \alpha \mathbb{I} + \beta J ; \qquad \alpha = 1-\frac{2}{N},\,\beta = \frac{2}{N^2},
\end{equation}
so that
\begin{equation}
    F-F^2 =  (\alpha- \alpha^2) \mathbb{I}+\left(\beta -\beta^2 N-2 \alpha \beta \right) J.
\end{equation}
Replacing this expression in Eq.(\ref{eq:aux1}),
we end up with
\begin{eqnarray}
\lim_{t \rightarrow \infty} \bar{K}(t) = \frac{1}{2}\left|\alpha-\alpha^2+\beta-\beta^2N-2\alpha \beta\right|
+(N-1)\left|\beta-\beta^2N-2\alpha\beta\right|,
\end{eqnarray}
which coincides with Eq.(\ref{eq:kbinf}).

\subsection*{Cycle graph}\label{ap:cycle graph}
For the cycle graph, we can indeed still use Eqs.(\ref{eq:aux0}) and (\ref{eq:aux1}),
but now the projectors refer to the eigenvalues
\begin{equation}
    \lambda_k = 2 - 2 \cos\!\left(\frac{2\pi k}{N}\right), \qquad k = 0,1,\dots,N-1,
\end{equation}
where $\lambda_0$ is nondegenerate and, if $N$ is even, also $\lambda_{N/2}$ is nondegenerate, while for
$k=1,\dots,\big\lfloor\frac{N-1}{2}\big\rfloor$ one has $\lambda_k=\lambda_{N-k}$,
and eigenvectors
\begin{equation}
    \ket{\psi_k} = \frac{1}{\sqrt{N}} \sum_{j=0}^{N-1} e^{i \frac{2\pi}{N}kj} \ket{j};
\end{equation}
the corresponding projectors are therefore
\begin{equation}
    \Xi_0=\ket{\psi_0}\!\bra{\psi_0},\qquad
\Xi_k=\ket{\psi_k}\!\bra{\psi_k}+\ket{\psi_{N-k}}\!\bra{\psi_{N-k}}\ \ (1\le k\le \lfloor\tfrac{N-1}{2}\rfloor),
\end{equation}
and, if $N$ is even,
$
\Xi_{N/2}=\ket{\psi_{N/2}}\!\bra{\psi_{N/2}}.
$

By translational invariance $F_{x,y}$ is a circulant matrix, i.e. it depends only on $d = (x-y)\bmod N$.
Using the explicit form of the projectors,
and introducing
$$
M = \lfloor\tfrac{N-1}{2}\rfloor,
$$
one finds
\begin{equation}
F_{x,y} = \frac{1}{N^2}\left( 1 + \sum_{k=1}^{M}
\left|e^{i\frac{2\pi}{N} k d}+e^{-i\frac{2\pi}{N} k d}\right|^2
+ \frac{1+(-1)^N}{2}\right)
= \frac{1}{N}+\frac{2}{N^2}\sum_{k=1}^{M}
\cos\!\Big(\frac{4\pi k}{N}\,d\Big),
\label{eq:F-cos-sum}
\end{equation}
and evaluating the finite trigonometric sum one gets 
\begin{equation}
F_{x,y} =
\begin{cases}
\displaystyle \dfrac{N-1}{N^2}, & N\,\text{odd} \ \text{and}\ d \neq0,\\[6pt]
\displaystyle \dfrac{N-2}{N^2}, & N\,\text{even} \ \text{and}\ d \neq0,\\[6pt]
\displaystyle \dfrac{2N-1}{N^2}, & N\,\text{odd} \ \text{and}\ d = 0,\\[6pt]
\displaystyle \dfrac{2(N-1)}{N^2}, & N\,\text{even} \ \text{and}\ d =0 \ \text{or}\ d= \frac{N}{2}.
\end{cases}
\end{equation}
From this, since also $F^2$ is a circulant matrix with
\begin{equation}
(F^2)_{x,0} = \sum_y F_{y,0}F_{x,y},
\end{equation}
we get
\begin{equation}
(F-F^2)_{x,y} =
\begin{cases}
\displaystyle -\dfrac{N-1}{N^3}, & N\,\text{odd} \ \text{and}\ d \neq0,\\[6pt]
\displaystyle -\dfrac{2(N-2)}{N^3}, & N\,\text{even} \ \text{and}\ d \neq0,\\[6pt]
\displaystyle \dfrac{(N-1)^2}{N^3}, & N\,\text{odd} \ \text{and}\ d = 0,\\[6pt]
\displaystyle \dfrac{(N-2)^2}{N^3}, & N\,\text{even} \ \text{and}\ d =0 \ \text{or}\ d= \frac{N}{2}.
\end{cases}
\end{equation}
Inserting this into Eq.~\eqref{eq:aux1}, we finally arrive at (compare with Eq.(\ref{eq:asnn})):
\begin{eqnarray}
\lim_{t\to\infty}\bar{K}(t)=
\begin{cases}
\displaystyle \frac12\left[ \frac{(N-1)^2}{N^3} + (N-1)\frac{(N-1)}{N^3} \right] = \frac{(N-1)^2}{N^3}, & N\,\text{odd},\\[6pt]
\displaystyle \frac12\left[ 2\frac{(N-2)^2}{N^3} + (N-2)\frac{2(N-2)}{N^3} \right]
= \frac{2(N-2)^2}{N^3}, & N\,\text{even}.
\end{cases}
\end{eqnarray}

\subsection*{A nondegenerate graph with $\lim\limits_{t \to \infty} \bar{K}(t) >0$}
\label{app:nondegenerate_example}

In this subsection we provide an explicit example showing that spectral
degeneracies of the Laplacian are not necessary for a nonzero asymptotic
time-averaged Kolmogorov quantifier under energy-basis dephasing (intrinsic
decoherence), for site-localized initial states and site measurements.

Consider the line graph $P_3$ (three sites in a line $1\!-\!2\!-\!3$), with
Laplacian
\begin{equation}
L=\begin{pmatrix}
1 & -1 & 0\\
-1 & 2 & -1\\
0 & -1 & 1
\end{pmatrix};
\end{equation}
the eigenvalues and eigenvectors of $L$ are
\begin{align}
\lambda=0:&\quad |\phi_0\rangle=\frac{1}{\sqrt3}(1,1,1)^{\mathsf T},\\
\lambda=1:&\quad |\phi_1\rangle=\frac{1}{\sqrt2}(1,0,-1)^{\mathsf T},\\
\lambda=3:&\quad |\phi_3\rangle=\frac{1}{\sqrt6}(1,-2,1)^{\mathsf T}.
\end{align}
Since the spectrum is nondegenerate, the corresponding projectors are $\Xi_\lambda=|\phi_\lambda\rangle\langle
\phi_\lambda|$ and thus -- see Eq.(\ref{eq:aux0}) --
\begin{equation}
F_{x,y}=\sum_{\lambda\in\{0,1,3\}} |\phi_\lambda(x)|^2\,|\phi_\lambda(y)|^2.
\end{equation}
A direct calculation yields
\begin{equation}
F=
\begin{pmatrix}
\frac{7}{18} & \frac{2}{9} & \frac{7}{18}\\[4pt]
\frac{2}{9} & \frac{5}{9} & \frac{2}{9}\\[4pt]
\frac{7}{18} & \frac{2}{9} & \frac{7}{18}
\end{pmatrix},
\qquad
F-F^2=
\begin{pmatrix}
\frac{1}{27} & -\frac{2}{27} & \frac{1}{27}\\[4pt]
-\frac{2}{27} & \frac{4}{27} & -\frac{2}{27}\\[4pt]
\frac{1}{27} & -\frac{2}{27} & \frac{1}{27}
\end{pmatrix}.
\end{equation}
Therefore, from Eq.~\eqref{eq:aux1} one obtains that for $\nu=1$ or $\nu=3$,
$\lim\limits_{t \to \infty} \bar{K}(t) = \frac{2}{27}$,
while for $\nu=2$, $\lim\limits_{t \to \infty} \bar{K}(t) = \frac{4}{27}$:
the graph $P_3$ has a nondegenerate Laplacian spectrum, yet it
exhibits a strictly positive asymptotic time-averaged multi-time nonclassicality
 for site-localized initial conditions and site measurements
under energy dephasing.

\end{document}